\documentclass[dvips]{article}

\usepackage{icrc2011}
\usepackage{amsmath,amssymb}

\title{2nd-order Fermi acceleration as the origin of the Fermi bubbles}

\newcommand{\etal}{\MakeLowercase{\textit{et al. }}} 
\shorttitle{Mertsch \etal 2nd-order Fermi acceleration as the origin of the Fermi bubbles}

\authors{Philipp Mertsch, Subir Sarkar}
\afiliations{Rudolf Peierls Centre for Theoretical Physics, University of Oxford, Oxford OX1 3NP, UK}
\email{p.mertsch1@physics.ox.ac.uk}

\abstract{Gamma-ray data from Fermi-LAT show a bi-lobular structure
  extending up to 50 degrees above and below the Galactic centre,
  coincident with a possibly related structure in the ROSAT X-ray map
  which presumably originated in some energy release close to the
  centre a few million years ago. It has been argued that the
  gamma-rays arise due to inverse Compton scattering of relativistic
  electrons accelerated at plasma shocks present in the bubbles. We
  explore the alternative possibility that the relativistic electrons
  undergo stochastic 2nd-order Fermi acceleration in the entire volume
  of the bubbles by plasma wave turbulence. This turbulence is
  generated behind the outer shock and propagates into the bubble
  volume, leading to a non-trivial spatial variation of the electron
  spectral index. Rather than a constant volume emissivity as
  predicted in other models we find an almost constant surface
  brightness in gamma-rays and also reproduce the observed sharp edges
  of the bubbles. We comment on possible cross-checks in other
  channels.}

\begin{document}
\maketitle

\section{Introduction}

Recently, data from the Fermi-LAT have revealed~\cite{Su:2010qj,Fermi}
(see also \cite{Dobler:2009xz}) the presence of two huge bi-lobular
structures in gamma-rays, the so-called ``Fermi bubbles'', extending
up to $50^\circ$ above and below the galactic plane. The overall
spectrum of the bubbles is $\propto E^{-2}$, i.e. much harder than the
$\pi^0$, inverse Compton (IC) and bremsstrahlung foregrounds from
galactic cosmic rays in the disk, and extends from a spectral shoulder
at about a GeV up to a cut-off/roll-over at a few hundreds of GeV. The
bubbles have an almost constant surface brightness with sharp
edges. The above properties as well as the size and position of the
bubbles are rather robust with respect to the details of foreground
subtraction making it unlikely that the bubbles are an artefact of the
foreground subtraction.

Both the position at galactic longitude $\ell = 0^\circ$ and its
symmetry with respect to the galactic plane hint at the galactic
centre (GC) as the origin of the bubbles. While similar structures
have been observed in radio galaxies the detection of the Fermi
bubbles is puzzling given that there is no evidence for present
activity of the massive black hole at the GC.  Understanding this
would provide an excellent probe of this region which is otherwise
obscured by the galactic disk.  The bubbles may play an important role
in the dynamics of our galaxy and constitute a source of cosmic
rays. They dominate the high latitude $\gamma$-ray emission at (and
possibly contribute close to) the GC, so constitute an important
background for indirect dark matter searches. It is therefore
important to understand and model the origin of the non-thermal
emission from the bubbles.

The observed high-energy gamma-rays can in principle be of hadronic or
leptonic origin, i.e. $\pi^0$ decay or inverse Compton (IC) scattering
of the ambient radiation fields (CMB, far IR and optical/UV) by high
energy electrons. A model~\cite{Crocker:2010dg} for the hadronic
origin invokes an increased star formation rate close to the GC in
combination with a strong convective wind up to kiloparsec distances
from the galactic plane. The neutral pions are produced from
spallation on the ambient gas in the bubbles of an $E^{-2}$ spectrum
of protons and nuclei accelerated in supernova remnants close to the
GC; the kinematics of the $\pi^0$ production can explain the spectral
shoulder around $1 \, \mathrm{GeV}$. Leptonic models, on the other
hand, have to explain how despite the rapid cooling of electrons the
bubbles volume can be filled with a consistently hard spectrum. It has
been argued that disruption of stars close by the central massive
black hole can heat the ambient gas and produce shocks. Estimates for
the rate of this process predict hundreds of concentric shock fronts
filling the bubbles. Electrons are repeatedly accelerated by diffusive
shock acceleration, thereby explaining the hard gamma-ray
spectrum. The spectral cut-off/roll-over would then be due to
competition of acceleration and cooling by IC (and possibly
synchrotron) losses.

Data from the ROSAT x-ray satellite~\cite{Snowden:1997ze} however only
show evidence for a limb brightened structure coinciding with the
bubble edges, possibly from a shock front. The non-observation of
x-rays from the bubble interior, on the other hand, points at a
relatively thin, hot plasma. With estimates for a gas density of $n
\sim 10^{-2} \, \mathrm{cm}^{-3}$ and a temperature of $T \sim 2 \,
\mathrm{keV}$, the total energy in hot gas is \mbox{$\sim 10^{54-55}\,
  \mathrm{erg}$}~\cite{Su:2010qj}. Furthermore, assuming velocities
typical for shock fronts in the interstellar medium gives $\sim 10^7
(U/1000\,\text{km} \, \mathrm{s}^{-1}) \mathrm{yr}$ for the age of the
bubbles at a projected distance of $10 \, \mathrm{kpc}$.

Other possible scenarios for the generation of the bubbles than those
above include jets emanating from the central black hole. While we
choose to remain agnostic about the origin of the bubbles itself we
note that a shock might have been produced by such a jet active for a
few million years. It has recently been shown~\cite{Guo:2011eg} that a
light but overpressured jet powered by $\sim 10 \, \%$ of the
Eddington luminosity leads to a shock coincident with the bubble edge
and in agreement with the overall bubble shape. In the following, we
will explain the non-thermal emission from the bubbles by 2nd-order
Fermi acceleration of electrons and IC scattering of these electron on
ambient radiation fields. As the electrons are constantly accelerated
in the whole bubble this can lead to the hard gamma-ray spectrum.

\section{Second order Fermi acceleration}

In particular, we start from the evidence for a shock front from
ROSAT. At the outer shock Rayleigh-Taylor and Kelvin-Helmholtz
instabilities will generate plasma turbulence that is then being
convected into the bubble interior by the downstream bulk flow. The
turbulence will cascade from the injection scale $L$ to smaller scales
and will finally be dissipated at a scale $l_d$, i.e. once the eddy
velocity reaches the Alfv\'en velocity, $v_\mathrm{edd}(l_d) \approx
v_A$. The usual Rankine-Hugoniot conditions allow to
compute~\cite{Fan:2009kr} the spatial variation of the eddy velocity
at the injection scale, $u$, and the magnetosonic phase velocity, $v_
\mathrm{F}$, with distance $x = \xi L$ from the shock:
\begin{eqnarray}
u(\xi) &=& \frac{U}{4} \frac{1}{C_1 \xi/3 + a^{-1/2}}\,, \label{eqn:u} \\
v_\mathrm{F}(\xi) &=& \frac{U}{4} \left( {5 - \frac{5}{3(C_1 \xi/3)^2} 
 + 4 \frac{v_\mathrm{A}^2}{U^2}} \right)^{1/2} \,, \label{eqn:vF}
\end{eqnarray}
where $U$ is the shock velocity, $v_\mathrm{A}$ the Alfv\'en velocity
(which we assume to be constant and equal to the speed of sound
$v_{\mathrm{s}, 0}$ at the shock) and $a = 3 - 16 v_{\mathrm{s}, 0}^2 /
U^2$.

\begin{figure}[!t]
  \vspace{5mm}
  \centering
  \includegraphics[width=\columnwidth]{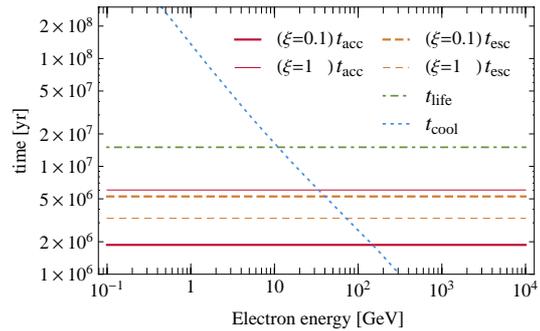}
  \caption{Relevant timescales as a function of energy. The
    acceleration and escape time depend on the distance $\xi = x / L$
    from the shock and are shown close to the shock ($\xi = 0.1$) and
    in the bubble interior ($\xi = 1$).}
  \label{fig1}
 \end{figure}

We consider the stochastic acceleration by large-scale, fast mode
turbulence~\cite{Ptuskin:1988aa}. Second order Fermi acceleration
processes like this have been proven successful in explaining the
non-thermal spectra of high-energy electrons in a variety of
astrophysical
environments~\cite{Scott:1975aa,Lacombe:1977aa,Achterberg:1979aa,Eilek:1979aa,Cowsik:1984aa,Fan:2009kr}
and might be responsible for the acceleration of ultra-high energy
cosmic
rays~\cite{Hardcastle:2008jw,Fraschetti:2008uc,O'Sullivan:2009sc}. The
spectrum is governed by the Fokker-Planck equation,
\begin{equation}
 \frac{\partial n}{\partial t} - \frac{\partial}{\partial p} 
 \left( p^2 D_{pp} \frac{\partial}{\partial p} \frac{n}{p^2} \right)  
 - \frac{n}{t_{\mathrm{esc}}} + \frac{\partial}{\partial p} 
 \left( \frac{\mathrm{d}p}{\mathrm{d}t} n \right) = 0 \, ,
\label{eqn:FokkerPlanck}
\end{equation}
where $n(p, t) \, {\rm d} p$ is the density of electrons with momentum
in $[ p, p + {\rm d} p]$. The second, third and fourth term describe
diffusion and systematic gains in momentum, escape due to spatial
diffusion and energy losses by synchrotron radiation and IC scattering
(with cooling time $t_\mathrm{cool} \sim p/({\rm d} p/{\rm d} t)$),
respectively. The diffusion coefficient in momentum for scattering by
fast magnetosonic waves is ~\cite{Ptuskin:1988aa}
\begin{equation}
D_{pp} = p^2 \frac{8 \pi D_{xx}}{9} \int_{1/L}^{k_\mathrm{d}}\mathrm{d} k \, 
 \frac{W(k) k^4}{v_\mathrm{F}^2 + D_{xx}^2 k^2}\, ,
\end{equation}
which translates into the timescale for acceleration, $t_\mathrm{acc}
\sim p^2/D_{pp}$. Both $t_\mathrm{acc}$ and $t_\mathrm{esc}$ (and
therefore the resulting spectrum) depend on three parameters that
cannot be inferred directly from observations. The scale of turbulence
injection $L$ is necessarily smaller than the size of the bubbles and
MHD simulations show generation of turbulence on kiloparsec
scales. Here, we assume $L = 2 \, \mathrm{kpc}$. The shock velocity
can in principle be determined from the displacement of the shock; the
shock needs $\sim
50\,(U/10^8\,\mathrm{cm}\,\mathrm{s}^{-1})\,\mathrm{yr}$ to move a
distance corresponding to the $1''$ resolution of the Chandra X-ray
observatory. Here, we fix $U = 2.6 \times 10^8
\,\mathrm{cm}\,\mathrm{s}^{-1}$, a value consistent with MHD
simulations~\cite{Guo:2011eg}. Finally the Alfv\'en velocity is given
by the square root of the ratio of magnetic field energy density to
thermal plasma energy density: $\beta_\mathrm{A} = v_\mathrm{A}/c =
\sqrt{U_B/U_\rho}$. Hence $\beta_\mathrm{A} > 2.8 \times 10^{-4}$ for
an estimated upper limit on the thermal gas density $n \lesssim
10^{-2}\,\mathrm{cm}^{-3}$~\cite{Su:2010qj} and a magnetic field $B =
4 \, \mu \mathrm{G}$. Such a field strength in the halo is suggested
by radio observations of edge-on spiral galaxies as NGC
891~\cite{Beck:1979aa}.  Here we adopt $\beta_\mathrm{A} = 5 \times
10^{-4}$. The resulting timescales are compared in Fig.~\ref{fig1}.

With these adopted parameters the dissipation length \mbox{$l_\text{d}
  > 8 \times 10^{19} (L/{\text{kpc}}) (U/{10^8 \, \text{cm} \,
    \text{s}^{-1}})^{-3} (\beta_\text{A}/{10^{-3}})^3\,\text{cm}$} is
always larger than the gyro-radius of relativistic electrons $\sim 7.5
\times 10^{11} (B / {4 \, \mu \text{G}})^{-1} (E/
\text{GeV})\,\text{cm}$. Hence, the spatial diffusion coefficient and
also the escape and acceleration time are effectively
energy-independent. Furthermore, with the parameters as above we
recover a hierarchy of timescales, $t_\mathrm{acc}, t_\mathrm{esc} \ll
t_\mathrm{life}$ which justifies the use of the steady state
solution~\cite{Stawarz:2008sp},

\begin{equation}
n(p) \propto \left \{
\begin{array}{ll}
p^{-\sigma} & \quad \mathrm{for} \quad p \ll p_{\mathrm{eq}} \, , \\
p^2 {\rm e}^{-p/p_{\mathrm{eq}}} & \quad \mathrm{for} \quad p \sim p_{\mathrm{eq}} \,.
\end{array}
\right.
\end{equation}
The spectral index, $-\sigma = 1/2 - \sqrt{9/4 + t_\mathrm{acc}/t_
  \mathrm{esc}}$, is determined by the ratio of acceleration and
escape times alone and asymptotically approaches $-1$ as $t_
\mathrm{acc}/t_ \mathrm{esc} \rightarrow 0$.  One could argue that for
low energies the cooling time becomes larger than the dynamical time
scale $t_\mathrm{life}$ and that therefore the use of the steady state
solution is not strictly justified. However, it has been
shown~\cite{Becker:2006nz} for 2nd-order Fermi acceleration that
irrespective of the cooling rate the spectrum always attains the
steady state spectrum in a few times $t_\mathrm{acc}$ and that the
steady state solution can therefore be applied as long as
$t_\mathrm{acc} \ll t_\mathrm{life}$. The change in timescales
$t_\mathrm{acc}$ and $t_\mathrm{esc}$ with distance from the shock
front $\xi$ (through $u(\xi)$ and $v_\mathrm{F}(\xi)$) is therefore
adiabatic such that the electron spectrum relaxes quickly to its
steady state value.

The emissivity in gamma-rays is calculated in the most general
form~\cite{Blumenthal:1970gc} using a recent model of interstellar
radiation fields~\cite{Porter:2005qx} and depends on the distance from
the shock. We calculate the flux of photons by integrating along the
line of sight through the bubble. Since the overall normalisation of
the electron spectrum depends on the microphysics of injection, we fix
this by demanding that our model matches the observed gamma-ray flux.

\section{Results}

\begin{figure}[!b]
  \vspace{5mm}
  \centering
  \includegraphics[width=\columnwidth]{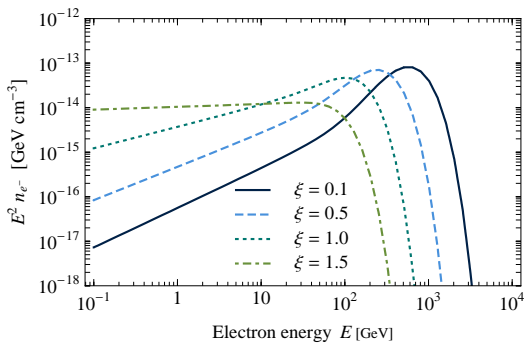}
  \caption{The electron spectrum $E^2 n_{e^-}$ at different distances $\xi = x / L$ from the shock front.}
  \label{fig2}
 \end{figure}

Figure~\ref{fig2} shows the electron spectrum for different distances
from the shock. The spectrum is hardest close to the shock and becomes
gradually softer towards the bubble interior. Furthermore, the
spectral pile-up and cut-off (determined by the competition between
acceleration and energy losses) move to lower energies. Integration of
the (position-dependent) spectrum over both bubbles shows that the
total energy in electrons above $100 \, \mathrm{MeV}$ is $\sim 10^{51}
\, \mathrm{erg}$. This is a rather moderate energy demand, in
particular in comparison to the hadronic model~\cite{Crocker:2010dg}
which requires up to five orders of magnitude more energy in high
energy protons.

\begin{figure}[!t]
  \vspace{5mm}
  \centering
  \includegraphics[width= \columnwidth]{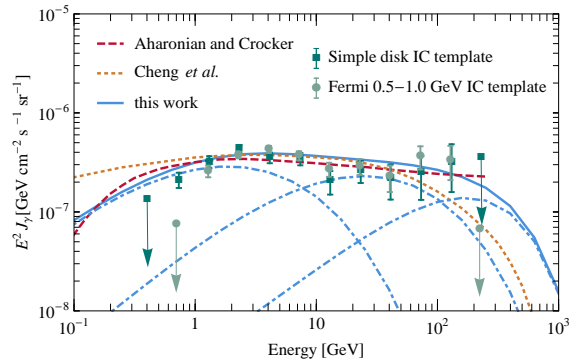}
  \caption{The overall spectrum $E^2 J_\gamma$ in gamma-rays. The data
    are shown as obtained with two different IC
    templates~\cite{Su:2010qj}. The fits from a
    hadronic~\cite{Crocker:2010dg} and a leptonic~\cite{Cheng:2011xd}
    model are shown by the dashed and dotted lines, respectively. The
    gamma-ray flux from our model is shown by the solid line and the
    dot-dashed lines show the contributions from IC scattering on the
    CMB, FIR and optical/UV (from left to right).}
  \label{fig3}
 \end{figure}

The overall spectrum of gamma-rays from the bubbles is shown in
Fig.~\ref{fig3} and compared to the data and predictions from other
models. Our model not only reproduces the $E^{-2}$ spectrum but also
both the spectral shoulder around $1 \, \mathrm{MeV}$ and the
roll-over/cut-off at $\sim 200 \, \mathrm{GeV}$. We note that the
other two models presented would need to invoke a somewhat unmotivated
break in the proton/electron spectrum to produce \textit{both}
features whereas in our model they arise naturally due to the very
hard electron spectrum and the cut-off due to cooling.

\begin{figure}[!bt]
  \vspace{5mm}
  \centering
  \includegraphics[width= \columnwidth]{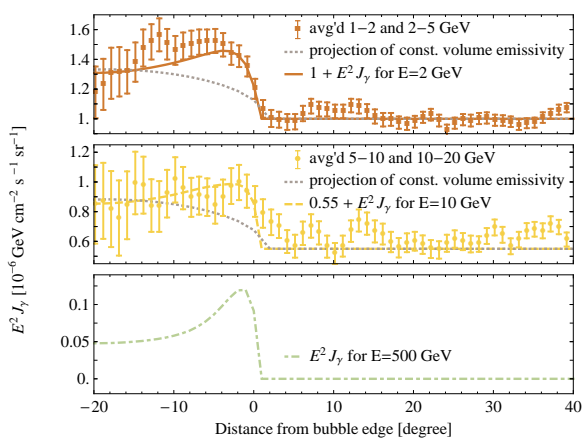}
  \caption{The intensity $E^2 J_\gamma$ in gamma-rays is shown as a
    function of the distance from the bubble edge at $2 \,
    \mathrm{GeV}$ (solid line), $10 \, \mathrm{GeV}$ (dashed line) and
    $500 \, \mathrm{GeV}$ (dot-dashed line), together with the
    data~\cite{Su:2010qj} from the averaged $1 - 2$ and $2 - 5 \,
    \mathrm{GeV}$ and the averaged $5 - 10$ and $10 - 20 \,
    \mathrm{GeV}$ maps. We also show the profile expected from a
    constant volume emissivity (dotted line) which clearly does not
    reproduce the observed profile.}
  \label{fig4}
 \end{figure}

In Fig.~\ref{fig4}, we compare the data with the intensity as a
function of distance from the bubble edge predicted by our model and
obtained in the same fashion as in Ref.~\cite{Su:2010qj},
i.e. averaging over great circles intersecting the bubble centre. At
both energies for which data is available ($2$ and $10 \,
\mathrm{GeV}$) our model nicely reproduces the constant profile inside
the bubbles and their sharp edges, i.e. the jump in intensity within a
few degrees around the bubble edge. We have also computed the profile
at $500 \, \mathrm{GeV}$ which is much more limb-brightened -- a
robust prediction of our model. We note that the profile expected from
a constant volume emissivity, as predicted by a
hadronic~\cite{Crocker:2010dg} and a leptonic~\cite{Cheng:2011xd}
model, would be much softer at the edges and does not reproduce the
data.
 
Both the constant surface brightness with sharp edges at low energies
and the limb-brightening at the highest energies are consequences of
the position dependent electron spectrum. As shown already in
Fig.~\ref{fig2}, the high-energy electrons are present only close to
the shock front whereas low-energy electrons are distributed over the
whole bubble volume. The IC emissivity at a given distance from the
shock is in fact a convolution over a certain range in electron
energies: For GeV gamma-rays the emissivity profile is flat in the
bubble interior and peaks at the bubble edge, which in projection
leads to a flat surface brightness with sharp edges. At the highest
energies however, only electrons of hundreds of GeV which are located
close to the shock can contribute which leads to a limb-brightening in
projection.

\begin{figure}[!bh]
  \vspace{5mm}
  \centering
  \includegraphics[width= \columnwidth]{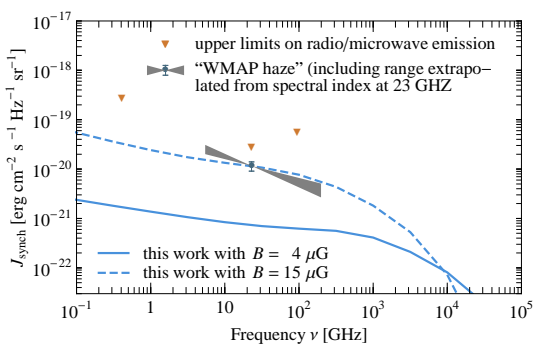}
  \caption{Radio and microwave flux from synchrotron emission of
    electrons in the bubble for a magnetic field of $4 \, \mu
    \mathrm{G}$ (solid line) and $15 \, \mu \mathrm{G}$ (dashed
    line). The data point shows the amplitude of the ``WMAP
    haze''~\cite{Dobler:2007wv} together with a range of spectral
    indices and the inverse triangles are upper limits obtained from
    the $408 \, \mathrm{MHz}$ all-sky survey~\cite{Haslam:1982zz} and
    the WMAP $23 \, \mathrm{GHz}$, $33 \, \mathrm{GHz}$
    bands~\cite{Jarosik:2010iu}.}
  \label{fig5}
 \end{figure}

While the ``WMAP haze'' \cite{Finkbeiner:2004us} has not been observed
in polarised emission~\cite{Gold:2010fm} and may be just an artefact
of the template subtraction~\cite{Mertsch:2010ga}, it has been
proposed as a physical counterpart of the Fermi
bubbles~\cite{Su:2010qj}. However as seen in Fig.~\ref{fig5}, in our
model the expected synchrotron flux in the middle of the bubble is of
the required amplitude \emph{only} if the magnetic field is as strong
as $15\,\mu \text{G}$. For a $4\,\mu \text{G}$ field the synchrotron
flux is significantly lower, \mbox{$1.6 \times 10^{-21} (\nu /
  \text{GHz})^{-0.2}\,\text{erg}\,\text{cm}^{-2}\,\text{s}^{-1}\,\text{sr}^{-1}$}.

The hadronic model predicts a detectable flux of neutrinos for the
proposed Mediterranean km$^3$ neutrino
telescope~\cite{Crocker:2010dg}. We stress however that the observed
bubble profile already disfavours this model (as well as the leptonic
DSA model) and instead favours our model with 2nd-order Fermi
acceleration of electrons.

\clearpage


\begin{thebibliography}{}

\bibitem{Su:2010qj}
  M.~Su, T.~R.~Slatyer, D.~P.~Finkbeiner,
  Astrophys.\ J.\, 2010, {\bf 724}: 1044

\bibitem{Fermi}
Fermi: http://www.nasa.gov/fermi.

\bibitem{Dobler:2009xz}
  G.~Dobler {\it et al.}, 
  Astrophys.\ J.\, 2010, {\bf 717}: 825

\bibitem{Crocker:2010dg}
  R.~M.~Crocker, F.~Aharonian,
  Phys.\ Rev.\ Lett.\, 2011, {\bf 106}:101102.

\bibitem{Snowden:1997ze}
  S.~L.~Snowden {\it et al.}, 
  Astrophys.\ J.\, 1997, {\bf 485}: 125
  
\bibitem{Guo:2011eg}
  F.~Guo, W.~G.~Mathews,
  [arXiv:1103.0055].

\bibitem{Fan:2009kr}
  Z.~Fan, S.~Liu, C.~L.~Fryer,
  Mon.\ Not.\ R.\ Astron.\ Soc.\, 2009, {\bf 406}: 1337.

\bibitem{Ptuskin:1988aa}
V.~S.~Ptuskin,
Sov.\ Astron\ Lett.\, 1988, {\bf 14}: 255.

\bibitem{Scott:1975aa}
J.~S. Scott, R.~A. Chevalier,
Astrophys. J. Lett. , 1975, {\bf 197}: L5.

\bibitem{Lacombe:1977aa}
C.\ Lacombe,
Astron.\ Astrophys.\, 1977, {\bf 54}: 1.

\bibitem{Achterberg:1979aa}
A.~Achterberg,
Astron.\ Astrophys.\, 1979, {\bf 76}: 276.

\bibitem{Eilek:1979aa}
J.~A.~Eilek, 
Astrophys.\ J.\  , 1979, {\bf 230}: 373.

\bibitem{Cowsik:1984aa}
R.~Cowsik, S.~Sarkar,
Mon.\ Not.\ R.\ Astron.\ Soc.\, 1984, {\bf 207}: 745.

\bibitem{Hardcastle:2008jw}
  M.~J.~Hardcastle {\it et al.}, 
  Mon.\ Not.\ R.\ Astron.\ Soc.\, 2009, {\bf 393}: 1041.

\bibitem{Fraschetti:2008uc}
  F.~Fraschetti, F.~Melia,
  Mon.\ Not.\ R.\ Astron.\ Soc.\, 2008, {\bf 391}: 1100.

\bibitem{O'Sullivan:2009sc}
  S.~O'Sullivan, B.~Reville, A.~M.~Taylor,
  Mon.\ Not.\ R.\ Astron.\ Soc.\, 2009, {\bf 400}: 248.

\bibitem{Beck:1979aa}
R.~Beck {\it et al.},
Astron.\ Astrophys.\ {\bf 77}, 25 (1979).

\bibitem{Stawarz:2008sp}
  L.~Stawarz, V.~Petrosian,
  Astrophys.\ J.\, 2008, {\bf 681}: 1725.

\bibitem{Becker:2006nz}
  P.~A.~Becker, T.~Le, C.~D.~Dermer,
  Astrophys.\ J.\, 2006, {\bf 647}: 539.

\bibitem{Blumenthal:1970gc}
  G.~R.~Blumenthal, R.~J.~Gould,
  Rev.\ Mod.\ Phys.\, 1970, {\bf 42}: 237.

\bibitem{Porter:2005qx}
  T.~A.~Porter, A.~W.~Strong,
  Proc. 29th Int. Cosmic Ray Conf., Pune,
  2005, {\bf 4}: 77.

\bibitem{Cheng:2011xd}
  K.~S.~Cheng {\em et al.}, 
  [arXiv:1103.1002].

\bibitem{Finkbeiner:2004us}
  D.~P.~Finkbeiner,
  [astro-ph/0409027].

\bibitem{Gold:2010fm}
  B.~Gold {\it et al.}, 
  Astrophys.\ J.\ Suppl.\, 2011, {\bf 192}: 15.

\bibitem{Mertsch:2010ga}
  P.~Mertsch, S.~Sarkar,
  JCAP, 2010, {\bf 1010}: 019.

\bibitem{Dobler:2007wv}
  G.~Dobler, D.~P.~Finkbeiner,
  Astrophys.\ J.\, 2008, {\bf 680}: 1222-1234

\bibitem{Haslam:1982zz}
  C.~G.~T.~Haslam, C.~J.~Salter, H.~Stoffel, W.~E.~Wilson,
  Astron.\ Astrophys.\ Suppl.\ Ser.\ , 1982, {\bf 47}: 1-142

\bibitem{Jarosik:2010iu}
  N.~Jarosik, C.~L.~Bennett, J.~Dunkley, B.~Gold, M.~R.~Greason, M.~Halpern, R.~S.~Hill, G.~Hinshaw {\it et al.},
  Astrophys.\ J.\ Suppl.\, 2011, {\bf 192}: 14

\end{thebibliography}
\end{document}